\documentclass[sigconf]{acmart}

\AtBeginDocument{%
  }

\setcopyright{acmlicensed}
\copyrightyear{2027}
\acmYear{2027}
\acmDOI{XXXXXXX.XXXXXXX}
\acmConference[SIGCSE TS 2027]{Proceedings of the 58th ACM Technical Symposium
on Computer Science Education}{February 17--20, 2027}{Sacramento, California, CA, USA}
\acmISBN{978-1-4503-XXXX-X/2018/06}

\usepackage{array}

\begin{document}

\title[When Does It Become Cheating?: Self-Efficacy, the Impostor Phenomenon, and Academic Dishonesty]{When Does It Become Cheating? Exploring the Roles of Self-Efficacy and the Impostor Phenomenon in Computing Students' Perceptions of Academic Dishonesty}

\author{Chandler C. Payne}
\email{cpayne67@gatech.edu}
\affiliation{%
  \institution{Georgia Institute of Technology}
  \city{Atlanta}
  \state{Georgia}
  \country{United States}}

\author{Jhonathan Sora-Cardenas}
\email{jcardenas45@gatech.edu}
\affiliation{%
  \institution{Georgia Institute of Technology}
  \city{Atlanta}
  \state{Georgia}
  \country{United States}}

\author{Olufisayo Omojokun}
\email{omojokun@cc.gatech.edu}
\affiliation{%
  \institution{Georgia Institute of Technology}
  \city{Atlanta}
  \state{Georgia}
  \country{United States}}

\author{Pedro Guillermo Feijóo-García}
\correspondingauthor
\email{pfeijoogarcia@gatech.edu}
\affiliation{%
  \institution{Georgia Institute of Technology}
  \city{Atlanta}
  \state{Georgia}
  \country{United States}}

\renewcommand{\shortauthors}{Payne et al.}

\begin{abstract}

Academic dishonesty is an urgent problem in computer science (CS) education, compounded by the rise of artificial intelligence (AI) and the prevalence of online learning in a post-COVID world. Self-efficacy and the impostor phenomenon are psychological constructs linked to academic dishonesty, yet little research has examined how they relate to how students perceive it. This study investigates the interplay between the impostor phenomenon, self-efficacy, and CS students' perceptions of academic integrity, in a context where online learning and AI agents have become commonplace. We conducted an online, asynchronous survey of undergraduate students (n = 443) in introductory to mid-level computing courses at a large Southeastern U.S. university. The survey collected measures of self-efficacy, responses to 13 potential cheating scenarios, the impostor phenomenon (Clance Impostor Phenomenon Scale), and demographic information. We observed a significant negative correlation between self-efficacy and the impostor phenomenon, and a positive correlation between the impostor phenomenon and judging an AI-related scenario as serious cheating. The impostor phenomenon was also correlated with perceived course dedication and prior teaching assistant (TA) experience, whereas self-efficacy was correlated with ethnicity. These findings highlight the need to better understand students' perceptions of AI and to keep the impostor phenomenon present in conversations about academic integrity in higher education scenarios.

\end{abstract}




\received{03 July 2026}
\received[accepted]{14 September 2026}

\maketitle

\section{Introduction}

Academic dishonesty remains an urgent problem in academia, especially in computer science (CS), where concern about academic misconduct is nearly ubiquitous among instructors \cite{wortzman_whos_2022}. Researchers have approached it from multiple angles, working to both understand and curb it through plagiarism detection in programming assignments \cite{albluwi_plagiarism_2020}, cheating-reduction strategies in introductory courses \cite{vahid_impact_2023}, and post-exam detection techniques \cite{salhofer_analysing_2017, xiao_detecting_2022}. Within engineering and computing disciplines, such misconduct has been attributed to three types of factors: demographic, psychological, and situational \cite{harding_current_2001}. These causes are not static, however. As campuses grow more diverse and technologically advanced, integrity education increasingly requires moving beyond established norms \cite{parnther_academic_2020}. These challenges have intensified as the educational context has shifted: the move to virtual assessments during COVID-19 raised cheating and plagiarism \cite{sozon_cheating_2024}, and the rise of artificial intelligence (AI) has compounded them further, introducing unclear integration guidelines, the trivialization of homework, and new avenues for cheating \cite{chassignol_artificial_2018, lee_teachers_2024, liu_leveraging_2023}.

The growing presence of AI in computing education raises questions not only about integrity, but also about its potential effects on psychological constructs relevant to learning, such as student self-efficacy and the impostor phenomenon. Self-efficacy refers to an individual's belief in their ability to perform the actions required to achieve specific outcomes \cite{bandura_self-efficacy_1977}, and has frequently been used to evaluate student success and proficiency \cite{schunk_self-efficacy_1985}. Self-efficacy has also been shown to be significantly related to academic misconduct among undergraduates. Peasah et al. \cite{peasah_role_2024} found that self-efficacy predicted academic fraud, while Ogilvie and Stewart \cite{ogilvie_integration_2010} found that academic self-efficacy predicted plagiarism intentions and moderated the effects of deterrence. Similarly, Baran and Jonason \cite{baran_academic_2020} found that general self-efficacy moderated the relationship between psychopathy and academic dishonesty. Together, these findings suggest that self-efficacy plays a meaningful role in students' susceptibility to academic dishonesty. Self-efficacy, however, does not operate in isolation; it is closely tied to students' broader psychological well-being.

Student well-being and mental health have become increasingly important concerns in computing education as workloads and pressure continue to build. One major mental health concern is the impostor phenomenon \cite{clance_impostor_1985}, described as internal feelings of inadequacy in one's abilities or intelligence despite objective evidence of successful performance \cite{clance_impostor_1985, menard_impostor_2023}. The impostor phenomenon is common among computing students, and understanding it gives insight into their mental health and well-being. It is higher in computing education than in many other areas of study \cite{rosenstein_identifying_2020}, and these feelings often persist as students enter major tech companies. Its relationship with academic misconduct appears counterintuitive: Ferrari \cite{ferrari_impostor_2005} found that non-impostors reported more plagiarism and cheating than impostors. Taken together, these findings suggest that understanding AI-related misconduct requires examining not only the behavior itself, but also the factors that lead a student to engage in it, placing self-efficacy and the impostor phenomenon at the center of this inquiry.

Despite evidence connecting both constructs to academic dishonesty, most prior work has focused on dishonest behavior itself rather than on how students perceive and judge it. Academic dishonesty in computing is defined and perceived differently by students, instructors, and administrators \cite{fraser_collaboration_2014}, and understanding these perceptions is crucial, as how students interpret potential misconduct shapes the choices they make. Yet there is limited research on how the impostor phenomenon relates to perceptions of academic dishonesty, particularly in the post-LLM/post-COVID context. This work therefore investigates the interplay between the impostor phenomenon and self-efficacy in computing students' perceptions of academic integrity. We examine how impostor-related self-doubt relates to self-efficacy and to how students interpret and justify integrity-related behaviors, particularly in new settings. Building on a previous study of students' perceptions of academic dishonesty \cite{chandler_arxiv, chandler_icer}, we examine this relationship in a post-COVID environment where online learning and large language model (LLM)-based agents have become commonplace. Thus, this exploratory study addresses the following research questions:

\noindent\textbf{RQ1:} How do undergraduate computing students respond to scenarios suggesting academic dishonesty in the CS classroom?

\noindent\textbf{RQ2:} How do self-efficacy and the impostor phenomenon affect undergraduate computing students' responses to scenarios suggesting academic dishonesty in the CS classroom?

\noindent\textbf{RQ3:} How does self-efficacy relate to the impostor phenomenon among undergraduate computing students in today's post-COVID educational landscape, where online learning and LLMs have become commonplace?

\noindent\textbf{RQ4:} How do undergraduate computing students' demographic factors relate to self-efficacy and the impostor phenomenon?

This work aims to fill a gap in existing research by connecting self-efficacy and the impostor phenomenon, to how undergraduate computing students approach possible situations of academic misconduct in the classroom.

\section{Background}

\begin{table*}[t]
\centering
\caption{Academic Dishonesty Scenarios Presented to Participants}
\label{tab:scenarios}
\small
\begin{tabular}{|>{\centering\arraybackslash}p{0.05\textwidth}|p{0.91\textwidth}|}
\hline
\textbf{ID} & \multicolumn{1}{c|}{\textbf{Academic Dishonesty Scenario}} \\
\hline

S1 &
You are working on a programming assignment. A classmate is having difficulty with their program. You look at their code to help identify the error. \\
\hline

S2 &
You are working on a programming assignment. A classmate is having difficulty with their program. You show the classmate a similar section of your code to help them understand. \\
\hline

S3 &
You are given an example of an already-compiled program that is executable. Your assignment is to create a program that runs like the example program. You decompile the example program and use parts of the resulting code in your assignment. \\
\hline

S4 &
At a review session, the TA goes over the types of questions you need to study for the exam. The TA gives example questions and answers them. When you receive the exam, you realize that the TA gave you the exact questions for the exam. Consider the TA's actions. \\
\hline

S5 &
You have spent three hours working on a portion of your assignment and you are having difficulty understanding it. There are resources (notes and/or assignments) from previous terms that answer your question. You look at those resources long enough to gain understanding. You have learned from those resources. You now use the information from those resources to finish your assignment. \\
\hline

S6 &
You use resources (notes and/or assignments) from a previous semester while studying for a quiz. When you take the quiz, it is identical to the resources that you found. You repeat all of the answers from the resources verbatim. Some of the answers are essay (open-ended) questions. \\
\hline

S7 &
You have an assignment due. However, you have not yet had time to complete it due to an overload of coursework. You get a time exemption by telling your professor that you have been sick. \\
\hline

S8 &
Your syllabus lists a website that you are allowed to use. You use an algorithm from this website without citing the source. \\
\hline

S9 &
In recitation one week, your TA goes over the type of questions you need to study for next week's quiz. Your TA gives example questions and then answers them. When you get the quiz the next week, you realize your TA gave the exact questions from the quiz. You write the exact answers you were given in recitation. \\
\hline

S10 &
Your professor has forbidden group work on a particularly difficult homework assignment. You work on the assignment with someone else from the class. \\
\hline

S11 &
A student writes their own code for an assignment but runs into an error. They provide the code to ChatGPT and ask for help debugging the code. \\
\hline

S12 &
You are in an online version of the class you are taking now. You have an open-note exam and use the internet during said exam. \\
\hline

S13 &
Trying to better understand the material, a student asks ChatGPT to explain a concept. In an effort to help visualize the learning process, ChatGPT provides a section of code explaining the concept, and the student then understands where they went wrong. They use a small portion of code from the provided ChatGPT answer to edit the code they were having trouble with. \\
\hline

\end{tabular}
\end{table*}

\subsection{Student Integrity and Cheating}

Student integrity and how to increase it has long been a focus of educational research \cite{albluwi_plagiarism_2020, vahid_impact_2023, salhofer_analysing_2017, xiao_detecting_2022, wortzman_whos_2022}. Much of this work has centered on cheating detection and consequential prevention, with techniques that have been adapted and made more robust as technology and classroom settings evolve \cite{salhofer_analysing_2017, xiao_detecting_2022, hung_examinator_2023, cui_answer_2024}. This increased detection and punishment has been shown to successfully lower the occurrence of cheating \cite{salhofer_analysing_2017}. Detection and punishment are not the only approach, however; rethinking the classroom setting to encourage students themselves to act with greater integrity has shown promising results as well, lowering cheating indicators such as time spent on assignments or similarity scores \cite{albluwi_plagiarism_2020, vahid_impact_2023}. In particular, student-instructor relationships are a highly effective and highly regarded aspect of integrity conversations, yet as Wortzman et al. note, they remain underutilized \cite{wortzman_whos_2022}.

\subsection{Student Well-Being, Mental Health, and the Impostor Phenomenon}

The impostor phenomenon is a major obstacle in computing education, where it appears more often than in many other fields \cite{rosenstein_identifying_2020, kolligian_jr_perceived_1991}. Its relationship with academic dishonesty, however, appears counterintuitive: Ferrari found that non-impostors reported more plagiarism and cheating than impostors \cite{ferrari_impostor_2005}. The impostor phenomenon has been linked to increased anxiety and depression \cite{fraenza_role_2016}, and these feelings often persist as students enter the computing industry \cite{rosenstein_identifying_2020}. It is also shaped by context: Abramson et al. found it influenced by major and gender \cite{fie_summer}, while Feijóo-García et al. reported in 2024 its connection to ethnicity, institutional culture, and background among students from higher education institutions \cite{fie_impostor}. Despite these links, little research has examined how the impostor phenomenon relates to students' perceptions of academic dishonesty. The impostor phenomenon also plays a role in student self-efficacy \cite{pakozdy_imposter_2024}.

\subsection{Self-Efficacy in Computing Education}

Self-efficacy has frequently been used to gauge students' learning outcomes, perceived success, and confidence \cite{schunk_self-efficacy_1985, wiggins_you_2017, zajacova_self-efficacy_2005, luszczynska_general_2005}. Its relationship with academic dishonesty, however, remains debated: in a study of college freshmen, self-esteem significantly predicted academic dishonesty, while self-efficacy did not \cite{ancheta_self-esteem_2024}. High self-efficacy positively affects motivation and pushes students toward more specialized fields \cite{schunk_self-efficacy_1985, zajacova_self-efficacy_2005, luszczynska_general_2005}, and Ahmad and Safaria link higher self-efficacy to improved academic performance \cite{safaria_effects_2013}. Self-efficacy in computing is also shaped by context, including cultural narratives about men versus women \cite{isman_how_2009} and students' prior programming background, as Govender and Basak found that a prior programming course drastically improves incoming students' self-efficacy \cite{govender_investigation_2015}. To support students across this range, Wiggins et al. propose guided experimentation for those with low self-efficacy and student control and social dialogue for those with high self-efficacy \cite{wiggins_you_2017}. Still, little research has examined how self-efficacy relates to students' perceptions of academic dishonesty.

\subsection{Artificial Intelligence in Computing Education}

AI has been transforming computing classrooms in both positive and negative ways. It has been leveraged for tutoring \cite{liu_teaching_2024} and even substitute teaching \cite{lee_teachers_2024}, but it has also introduced challenges such as confusing integration guidelines, the trivialization of homework, and new avenues for cheating \cite{liu_leveraging_2023, chassignol_artificial_2018}. Effective integration is essential, as customized and personalized curricula have led to improvements in learning quality \cite{chen_artificial_2020}. As AI becomes commonplace in computing education, understanding how students perceive its use, particularly in situations that blur the line between help and misconduct, is increasingly important. This motivates our study of how self-efficacy and the impostor phenomenon relate to students' perceptions of academic integrity in this new context.

\section{Methodology}

\subsection{Context and Recruitment}

This study was conducted online and asynchronously during the Spring 2026 semester, recruiting participants from a major Southeastern university in the United States. Eligible participants were undergraduate students enrolled in introductory to mid-level computing courses. The questionnaire was distributed by course instructors through the university's Learning Management System (LMS), and all communication was handled asynchronously via email. Participation required approximately 15 minutes to complete. The study received approval from the Institutional Review Board (IRB) prior to data collection.

\subsection{Participants}

A total of 508 responses were received. After removing 61 incomplete or duplicate submissions and 4 responses that failed attention checks based on a simple multiple choice question, the final analytical sample consisted of $n$ = 443 participants.

The sample represented a range of demographic backgrounds. In terms of gender, 261 participants identified as male, 171 as female, and four ($n$ = 4) as non-binary or other; seven ($n$ = 7) participants chose not to disclose their gender. Regarding ethnicity, most participants identified as Asian ($n$ = 255) or White, non-Hispanic/non-Latin American ($n$ = 118); additional identities included Hispanic/Latin American ($n$ = 38), Black/African American ($n$ = 42), Middle Eastern/North African ($n$ = 18), and Pacific Islander ($n$ = 3). These totals exceed the overall sample size because multiracial participants were able to select multiple categories.

All participants were undergraduates. The majority identified as domestic students ($n$ = 375), with a smaller portion identifying as international students ($n$ = 68). Most participants were in their first year ($n$ = 247), followed by second-year students ($n$ = 140); 56 participants were in their third year or beyond. A smaller number reported prior teaching assistant (TA) experience ($n$ = 21), and 103 participants reported prior industry experience related to computing, compared to 340 who reported none.

This study reports on exploratory quantitative design findings. For the scope of this paper, our analysis focuses on the quantitative data, from measuring self-efficacy, responses to 13 potential cheating scenarios, the impostor phenomenon (Clance Impostor Phenomenon Scale \cite{clance_impostor_1985}), and student demographic data.

\subsection{Measures / Instruments}

\subsubsection{Self-Efficacy Questions}
    
    Self-efficacy was measured using the General Self-Efficacy Scale (GSE) developed by Schwarzer and Jerusalem \cite{schwarzer1995generalized}. The scale consists of 10 items rated on a four-point scale (1 = ``not at all true", 2 = ``barely true", 3 = ``moderately true", 4 = ``exactly true"), with a sample item being ``I can always manage to solve difficult problems if I try hard enough." Item responses were summed to produce a composite self-efficacy score ranging from 10 to 40, where higher scores indicate greater self-efficacy.

\subsubsection{Impostor Phenomenon Questions}

    The impostor phenomenon was measured using the Clance Impostor Phenomenon Scale (CIPS) \cite{clance_impostor_1985}. The scale consists of 20 items rated on a five-point scale (1 = ``not at all true", 2 = ``rarely", 3 = ``sometimes", 4 = ``often", 5 = ``very true"). Item responses were summed to produce a composite score ranging from 20 to 100, where higher scores indicate more frequent and intense impostor feelings. In accordance with the instrument's usage terms, individual items are not reproduced here; the full instrument is available from the original author \cite{clance_impostor_1985}. 

\subsubsection{Cheating Scenarios}\label{scenarios}

    Perceptions of academic dishonesty were measured using a set of 13 scenarios adapted from Ozment et al. \cite{article}. For each scenario, participants selected one of three classifications: (1) Not Cheating, (2) Trivial Cheating, or (3) Serious Cheating. Scenarios described a range of situations relevant to computing coursework, covering topics such as peer collaboration, use of online or past class materials, and additional adapted questions about AI usage. For analysis, responses were coded as 1 = Not Cheating, 2 = Trivial Cheating, and 3 = Serious Cheating. Table \ref{tab:scenarios} presents the 13 scenarios used for this exploratory study.
    
\subsubsection{Demographic Questions}
    
    Following these questions, participants responded to a set of demographic questions. For the scope of this work, demographics are reported on all participants’ age, race, gender, ethnicity, and domestic status, as well as the following demographics: Students’ academic year, prior work experience, TA experience, current number of credit hours, and perceived hours needed to succeed in the current course. 

\subsection{Procedure}

    Data was collected during the Spring 2026 semester via an online, asynchronous questionnaire distributed through the university's LMS. Participants first completed an informed consent form and then proceeded through the measures in the following order: self-efficacy, cheating scenarios, impostor phenomenon, and demographics. Completion took approximately 15 minutes.

\section{Results}

For the analysis, descriptive and inferential statistics were employed to evaluate and determine significant relationships. Spearman's rho was used to determine the strength of the correlation. Given the exploratory nature of this analysis, no multiple-comparison correction was applied. Instead, unadjusted $p$-values are presented to highlight potential patterns, which should be interpreted as tentative and validated in future confirmatory studies.

\subsection{Perceptions of Academic Dishonesty}

Descriptive statistics were used to examine student responses to the 13 academic dishonesty scenarios we presented (see Section \ref{scenarios}). 

\begin{figure*}[t]
    \centering
    \includegraphics[width=0.6\linewidth, keepaspectratio]{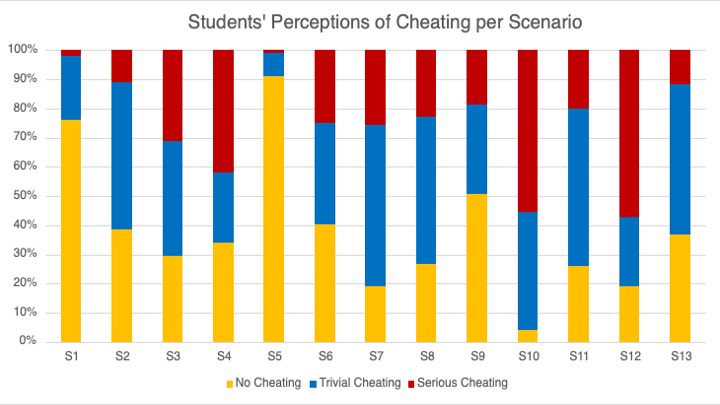}
    \caption{Student's Perceptions of Cheating per Scenario}
    \label{fig:placeholder}
\end{figure*}

Responding to \textit{RQ1}, students' responses varied considerably across the scenarios, suggesting that they distinguish between different forms and contexts of academic dishonesty rather than viewing cheating as a uniform concept (see Figure \ref{fig:placeholder}).

A notable pattern emerged when comparing scenarios involving assistance from other people (S4 and S10) with those involving AI-based assistance (S11 and S13). In the human-assisted scenarios, participants were more likely to classify the behavior as dishonest, frequently selecting \textit{``Serious Cheating."} In S4, the assistance came from a TA, whereas in S10 it came from a classmate. By contrast, responses to the AI-assisted scenarios more frequently classified the behavior as \textit{``Trivial Cheating."} This contrast may indicate that students conceptualize AI-based assistance differently from assistance provided by other people. Interestingly, S12, which described students seeking online support during an open-note exam, received a substantially higher proportion of \textit{``Serious Cheating''} responses. 

Finally, these scenario ratings served as the basis for subsequent analyses examining relationships with self-efficacy, impostor phenomenon, and demographic variables. No additional significant relationships among the scenario responses were observed beyond those reported in the analyses below.

\subsection{Self-Efficacy, Impostor Phenomenon, and Academic Dishonesty}

Responding to \textit{RQ2}, a highly significant positive correlation ($r_s = .133$, $p = .005$) was found between impostor phenomenon and S14 (see Table \ref{tab:scenarios}). This demonstrates that the higher feelings of impostor phenomenon there are, the more likely a student is to consider this scenario ``serious cheating.''

No significant relationships were found between self-efficacy and responses to the academic dishonesty scenarios. Likewise, no additional significant correlations were observed between impostor phenomenon and the remaining scenario questions.

\subsection{Self-Efficacy and Impostor Phenomenon Amid Online Learning and LLMs}

Responding to \textit{RQ3}, a significant negative correlation was found ($r_s = -.277$, $p < .001$) between self-efficacy and impostor phenomenon, indicating that the higher self-efficacy a student has, the less likely they are to experience impostor phenomenon. This aligns with our hypothesis as well as prior research. These results support a significant negative association between self-efficacy and the impostor phenomenon, consistent with prior literature indicating that greater confidence in one's abilities is linked to fewer impostor-related thoughts.

\subsection{Demographic Factors, Self-Efficacy, and the Impostor Phenomenon}

Responding to \textit{RQ4}, for demographic data on ethnicity and race, a one-hot encoding was performed prior to analysis to ensure proper tracking and representation for those that selected multiple races. For the White--non-Hispanic/non-Latin American ($n = 118$) and Asian ($n = 255$) populations, a significant relationship was found with self-efficacy, with the White--non-Hispanic/non-Latin American population seeing a positive correlation ($r_s = .161$, $p < .001$) and the Asian population seeing a negative correlation ($r_s = -.127$, $p = .008$). These were the two largest demographic groups, and it is interesting to see the higher reported self-efficacy in White students as opposed to Asian students.

Little to no further significant correlations were found with regard to race, ethnicity, gender, age, and domestic/international status. No additional significant demographic relationships with impostor phenomenon were observed.

Further, a positive correlation ($r_s = .114$, $p = .016$) was found between course dedication and impostor phenomenon. Course dedication refers to the amount of hours that a student considers is necessary to succeed in the course. The more hours they believed to be necessary, the more likely they were to experience impostor feelings. This is true for TA experience as well ($r_s = .106$, $p = .026$), where it was observed that the more experience a student has as a TA, the more impostor feelings they reported.

Number of credit hours, academic year, and prior work experience did not have any observable significant correlations. No additional significant demographic relationships with self-efficacy were observed. 

\section{Discussion}

Regarding RQ1, students' perceptions of academic dishonesty varied notably depending on whether the assistance came from another person or from AI, with the exception of S12. One possible explanation is that students perceived the assessment context (i.e., an exam) as inherently more restrictive than a regular assignment. Nevertheless, the observed differences suggest that students may apply more lenient judgments to AI-assisted behaviors, raising questions about whether they clearly distinguish between AI as a productivity tool, a learning aid, or a collaborative partner during academic work.

Turning to RQ2, one interesting finding is the relationship between impostor phenomenon and S13 of our survey (see Table \ref{tab:scenarios}). As one of only three questions in the survey related to AI, it interesting to see that this had such a high correlation to students' impostor feelings. With higher impostor feelings, it was observed that respondents were more likely to consider this situation serious cheating. It could be that the mindset of doubting one’s abilities leads to a higher likelihood of considering this situation academic misconduct, as if you were ``better” or more equipped to handle the situation you would not have to resort to this. Specifically, this situation references the idea of utilizing AI for help on homework, and utilizing the newly learned information to complete your assignment by using code provided by AI agents. Students experiencing the impostor phenomenon may believe they should be able to complete an assignment on their own and could interpret their inability to do so as a personal failure, making them more likely to consider this act as cheating.

The impostor phenomenon may play a particularly important role in this context, especially given that it has only recently begun to receive attention in education research and practice, particularly within computing education. Many students, teachers, and universities are still navigating how to interact with AI, and thus feelings of impostor phenomenon may be heightened in these spaces. Interestingly, the impostor phenomenon did not show signs of significant correlation with any other situations, which could evidence that new and unfamiliar situations can be grounds for heightened impostor feelings. 

Turning to RQ3, our findings are consistent with prior research showing an inverse relationship between self-efficacy and the impostor phenomenon. In our sample, students reporting higher self-efficacy were less likely to report impostor feelings. Although this relationship has been well established in prior work, observing it in our sample provides additional context for interpreting the relationships examined in this study.

Finally, turning to RQ4, a positive correlation was also observed between students' perceptions of required course dedication and the impostor phenomenon. The more hours reported as necessary to succeed in the class, the more likely students were to experience impostor phenomenon. Similarly, those that reported having more TA experience also demonstrated a higher experience of impostor phenomenon. It may be that this is due to the fact that TAs are often surrounded with high achieving individuals, namely professors and other TAs. Those who see the class as requiring more hours may also experience self-doubt about the amount of hours they put into the class. Further research should explore this phenomenon more thoroughly.

One of the most interesting findings was self-efficacy and the impact that ethnicity had on it. Those students that reported ethnic groups such as White--non-Hispanic/non-Latin American had significantly high self-efficacy, while those that indicated being Asian had significantly low self-efficacy. It is unsure why this would be the case in particular, but it is important to note the effect that ethnicity and race had on self-efficacy. No other ethnic groups were observed to have a significant difference in student self-efficacy or impostor phenomenon, which could mean that cultural differences may play a pivotal role here. Further research should explore this phenomenon in greater depth.

\vspace{-1 em}

\subsection{Limitations and Future Works}

One of the biggest initial limitations is the scope of the survey being limited to a single, rigorous university. Especially with respect to the impostor phenomenon, there may be a lack of generalization as students may be more prone to compare and contrast themselves to their high-achieving peers and struggle through difficult classes. Additionally, the demographic data of this work also leans heavily towards a White (non Hispanic or Latin American) and Asian population. This remains true with regards to age and college year as well, with a large majority of respondents being younger first or second year students with little to no internship or industry experience. 

There were a large number of respondents which may help to mitigate some of these limitations. Still, the scope of this work was limited and further research is highly encouraged to build upon this work. Furthermore, the exploratory attributes of this analysis warrant future research to affirm and validate the tendencies and relationships emphasized in this study. 

Given the highlighted relationship with AI and the impostor phenomenon, a possible new direction could be to study more specifically perceptions on AI integration, as well as how the impostor phenomenon can affect outlook on unfamiliar situations and contexts. Further studies may fully validate the findings of this study with further insight using multiple-comparison correction and fully evaluating the intricate, multi-dimensional relationships. 

Finally, Baker's \cite{baker_how_2026} account of a campus culture that rewards ``getting ahead at any cost" points to another valuable direction: extending this line of inquiry beyond the student population to the broader university ecosystem, including faculty, administrators, and staff, and examining how their perceptions and behaviors interact with this culture. Future research might also investigate how institutional policies relate to,  and potentially reinforce or mitigate, this culture, and identify which policy interventions are most effective in addressing it.

\section{Conclusion}

This exploratory study highlights the link between self-efficacy, impostor phenomenon, and student perceptions on academic misconduct. Evidence shows there is a strong link between impostor phenomenon and a higher likelihood to consider a situation cheating, especially with regard to situations pertaining the usage of AI. This indicates a need for better understanding of student AI perceptions, and better ways to inform and encourage students in the use of AI. This similarly motivates further study into the impostor phenomenon in recent developments and new contexts.

A significant initial correlation was also found between perceived course difficulty and the impostor phenomenon. It is hypothesized that this is due to the perceived need for more effort underscoring either a feeling of inadequacy in the amount of effort currently put in, or a self-doubt on intelligence when having to put in so many hours to succeed as compared to other students. In addition, a positive correlation was observed between TA experience and the impostor phenomenon as well, which could be attributed to a constant surrounding of high-achieving, intellectual peers and professors.

This work underscores the need for better understanding of student self-efficacy and impostor phenomenon, as it can have impact on areas of their academics that have not been explored before. Given that AI is positioned to continue to grow in relevance in the computing classroom, it's imperative that CS educators emphasize the importance of clear instructions on use and keeping topics like impostor phenomenon in conversations, as that can go a long way in helping students. There is much further research to be done to expand and verify these results, and understanding both the impact and contributing factors to self-efficacy and the impostor phenomenon will go a long way in helping students succeed well past higher education.

\begin{acks}
This material is based upon work supported by the National Science 
Foundation (NSF) under Grant No. 2434428. Any opinions, findings, or recommendations expressed in this material are those of the author(s) and do not necessarily reflect the views of the NSF.

The authors acknowledge the use of Grammarly, ChatGPT-5.6, and Claude Opus 4.8, AI-based language models developed by OpenAI and Anthropic, for spell-checking, grammar, and editing assistance. Tools were used to refine the language and enhance the manuscript's clarity. No content, figures, or images were generated by AI for this work.

\end{acks}

\bibliographystyle{ACM-Reference-Format}
\bibliography{references}
\end{document}